\documentclass[prl,english, reprint, aps,superscriptaddress,onecolumn]{revtex4-1}
\usepackage[T1]{fontenc}
\usepackage[utf8]{inputenc}
\usepackage{bm}
\usepackage{amsmath}
\usepackage{amssymb}
\usepackage{graphicx}
\usepackage{color}
\begin{document}
\title{Understanding the electromagnetic response of graphene/metallic nanostructures
hybrids of different dimensionality }
\author{ Tatiana G. Rappoport}
\affiliation{Department and Centre of Physics, and QuantaLab, University of Minho,
Campus of Gualtar, 4710-057, Braga, Portugal}
\affiliation{Instituto de F\'\i sica, Universidade Federal do Rio de Janeiro, Caixa
	Postal 68528, 21941-972 Rio de Janeiro RJ, Brazil}
\author{ Itai Epstein}
\affiliation{ICFO-Institut de Ciencies Fotoniques, The Barcelona Institute of Science and Technology, 08860 Castelldefels (Barcelona), Spain}
\author{Frank H. L. Koppens}
\affiliation{ICFO-Institut de Ciencies Fotoniques, The Barcelona Institute of Science and Technology, 08860 Castelldefels (Barcelona), Spain}
\affiliation{ICREA-Institució Catalana de Recerca i Estudis Avançats, Barcelona, Spain}
\author{ Nuno M. R. Peres}
\affiliation{Department and Centre of Physics, and QuantaLab, University of Minho,
Campus of Gualtar, 4710-057, Braga, Portugal}
\affiliation{International Iberian Nanotechnology Laboratory (INL), Av. Mestre
José Veiga, 4715-330, Braga, Portugal}

\date{\today}
\begin{abstract}
Plasmonic excitations such as surface-plasmon-polaritons (SPPs) and graphene-plasmons (GPs), carry large momenta and are thus able to confine electromagnetic fields to small dimensions. This property makes them ideal platforms for subwavelength optical control and manipulation at the nanoscale. The momenta of these plasmons are even further increased if a scheme of metal-insulator-metal and graphene-insulator-metal are used for SPPs and GPs, respectively. However, with such large momenta, their far-field excitation becomes challenging. In this work, we consider hybrids of graphene and metallic nanostructures and study the physical mechanisms behind the interaction of far-field light with the supported high momenta plasmon modes. While there are some similarities in the properties of GPs and SPPs, since both are of the plasmon-polariton type, their physical properties are also distinctly different. For GPs we find two different physical mechanism related to either GPs confined to isolated cavities, or large area collective grating couplers. Strikingly, we find that although the two systems are conceptually different, under specific conditions they can behave similarly. By applying the same study to SPPs, we find a different physical behavior, which fundamentally stems from the different dispersion relations of SPPs as compared to GPs. Furthermore, these hybrids produce large field enhancements that can also be electrically tuned and modulated making them the ideal candidates for a variety of plasmonic devices. \end{abstract}
\maketitle

Plasmons-polaritons are near-field evanescent modes, corresponding to the coupling of electromagnetic fields with collective oscillation of charge carriers~\cite{Maier2007,Goncalves2016}. These polaritons play a crucial role in nanophotonics due to their inherent ability to confine light at the nanoscale, owing to their capability to carry large momentum that is much higher than that of free-space photons. The two most widely used plasmons-polaritons in that aspect, are GPs for the mid-infrared (MIR) to Terahertz (THz) spectra, and SPPs for the visible (VIS) to near-infrared (NIR) spectra. These can carry extremely large momentum especially in the configuration of Metal-Insulator-Graphene (MIG) for GPs~\cite{Alonso-Gonzalez2017,Lundeberg2017,Iranzo2018,Goncalves2016,Goncalves2020} and Metal-Insulator-Metal (MIM) for SPPs~\cite{Bozhevolnyi2005,Miyazaki2006,Oulton2008, Lindquist2012,Moreau2012,Chen2013,Lassiter2013,Lassiter2014,Toma2014}.

The possibility of confining and manipulating MIR$\backslash$THz electromagnetic fields at nanoscopic scales can lead to a variety of novel applications in  photodetecting and sensing, especially for infrared molecular spectroscopy~\cite{Boriskina2013,Rodrigo2017,Hu2019}. From a technological perspective, it is crucial to sharply enhance and confine the electromagnetic field to increase the interaction between incident light and molecules, which leads to an increase in the vibrational absorption signals. The manipulation of fields at nanoscopic scales can also be used to reduce the footprint of MIR and THz devices~\cite{Woessner2017,Liu2015}.  

SPPs and localized-surface-plasmons (LSPs) in metallic nanoparticles and metallic gratings can squeeze light well below its diffraction limit, in the VIS$\backslash$NIR~\cite{Barnes2003, Moreau2012, Lassiter2013, Lassiter2014}. However, these have weak confinement in the MIR$\backslash$THz frequency range and limited tunability \cite{Halas2011}. Thus, a promising approach to overcome these limitations and achieve unprecedented level of nanoscale manipulation is the use of GPs. These exhibit high degree of confinement of the free-space wavelength, which is several orders of magnitude smaller in the frequency range of interest~\cite{Jablan2009,DeAbajo2014,Park2015,Koppens2014,Low2014,Goncalves2016,Celano2019}. 

The dispersion relation (DR) of both SPPs and GPs exhibit momenta that is larger than that of the free-space photon, namely the light-line~\cite{Goncalves2016}. However, the DR of SPPs lies quite close to the light-line, and thus disperse very little compared to GPs, which show considerable dispersion in all of the frequency-momentum phase-space \cite{Goncalves2016}. Both MIM and MIG configurations, involving a thin insulating layer, can hold a vertically confined mode with even larger momentum \cite{Goncalves2016}. However, the increase in the momentum of the MIM mode in accompanied with large losses. This is not the case for the MIG  mode, since the metal contribution to losses in this system is minimal \cite{Alonso-Gonzalez2017}, even though its momentum is much larger than that of the MIM case. At such high momentum, the MIG DR becomes linear and the mode is thus referred to as an acoustic graphene plasmon (AGP)\cite{Principi2011,Alonso-Gonzalez2017,Iranzo2018}.

Owing to the large momenta carried by these modes, and the fact that they are near-field evanescent modes, most experimental approaches, especially in the case of GPs, rely on near-field excitation of these plasmons \cite{Chen2012,Fei2012,Brar2013,Fang2013,Yan2013,Alonso-Gonzalez2017,Ni2018}. However, it is extremely important to understand their interaction with far-field light in order to enable the broad usage of these high momentum modes for light-matter interaction and nanophotonics applications. It is in this context, and motivated by recent experiments where highly confined AGPs are excited from the far-field with either periodic metallic rods~\cite{Iranzo2018,Lee2019} or randomly dispersed  nanocubes~\cite{Epstein2020}, we study the physical coupling mechanisms behind the two approaches. We explore the tunability of different resonator architectures, consisting of a single graphene layer, a spacer, and a silver nanostructure. We address the differences and similarities between AGPs excited by an isolated metallic structure and a periodical one, in both 2D and 3D configurations. We show that the electric and magnetic fields can be enhanced and manipulated differently for these various dimensions. In addition, and in the context of bridging the gap between the communities working in noble-metal plasmonics and in graphene plasmonics, we present a comparison between similar structures involving either the hybrid AGP structures on one side (MIG), and metallic films coupled to metallic nanoparticles (MIM). Finally, we show that for the case of AGPs excited with metallic nanocubes, one obtains sizeable dipolar magnetic resonances. These can be continuously tuned throughout the MIR spectrum by varying the graphene Fermi-level and the device geometry. 

\begin{figure}[h]
\centering{\includegraphics[width=0.95\columnwidth]{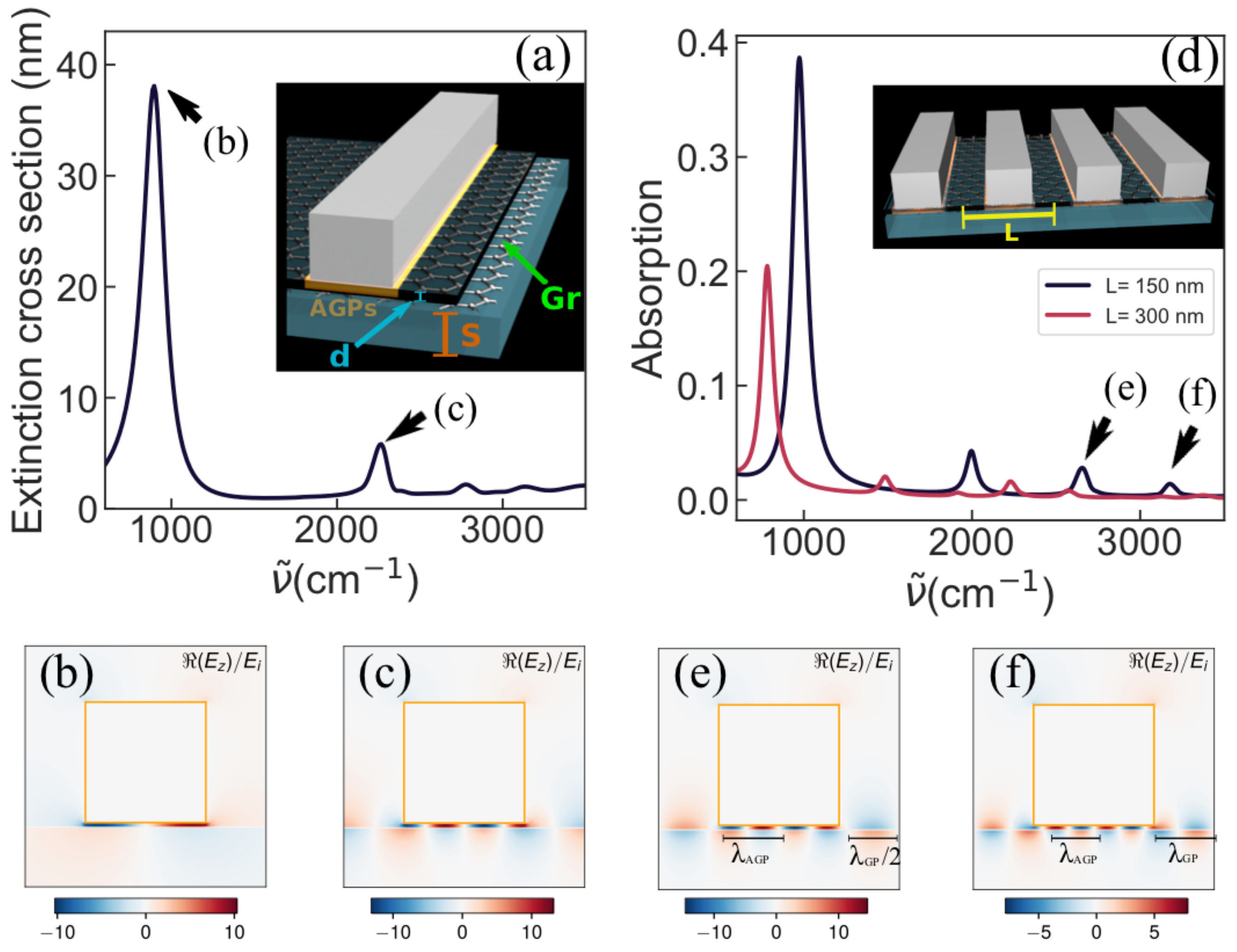}}
\caption{\label{fig1}(a) Extinction cross section spectra of an AGP resonator architecture consisting of a substrate $S$, a single graphene layer, a spacer $d$ and a silver rod with lateral cross-section $W$=75 nm as shown in the inset. (b)-(c) Electric field distributions showing the real component of $E_z(\vec{r})/E_i$  for the first and second modes respectively of the spectra in (a). (d) Absorption spectra of an architecture consisting of a substrate a single graphene layer, a spacer and periodic structure of silver rod  with period $L$ as shown in the inset. (e)-(f) Electric field distributions showing the real component of $E_z(\vec{r})/E_i$  for the third and fourth modes respectively of the spectra in (d).  The colors represent the spatial distributions (in plane $xz$) of the electric field component $E_z(\vec{r})$ for $E_F$=0.4 eV normalized by the intensity of incident fields $E_i$. }
\end{figure}

Our basic building block consists of a monolayer graphene on top a bulk substrate S with permittivity $\varepsilon_S$, separated from a metallic nanostructure by a thin dielectric spacer, of thickness $d$ and permittivity $\varepsilon_d$ (Fig. \ref{fig1}). A metallic nanoparticle on top of a graphene layer can be seen as a coupled plasmonic structure where the thickness of the dielectric layer controls the plasmonic coupling~\cite{Lassiter2013,Maurer2015}. We explore AGP confinement and electromagnetic field enhancement in four different types of designs of silver nanostructures, with different dimensionalities: 
 An isolated cube of width $W$, which is a zero-dimensional plasmonic structure,  an isolated rod with square cross-section of width $W$, which is a one-dimensional plasmonic structure, and periodic structures of rods and cubes with period $L$, which are 1D and 2D plasmonic crystals respectively. For this purpose, we perform full-wave finite element frequency domain simulations~\cite{comsol}. For simplicity, graphene is simulated as a single layer with optical conductivity at room temperature that is given by a Drude like expression $\sigma_g(\omega)=4\sigma_0 E_F/(\pi(\hbar\gamma-i\hbar\omega))$~\cite{Goncalves2016},  where $\sigma_0=e^2/2\hbar$, $E_F$ is the Fermi energy, $\gamma$ is the relaxation rate and $\omega$ is the frequency of the incident light. This approximation is known to provide qualitative agreement with the experiments. Quantitative agreement requires the use of graphene's nonlocal conductivity, which become especially important at small graphene doping and short graphene-metallic structure distances. If these two conditions are met, the position of the resonances is shifted relatively to that computed using the local approach, and the mode volume saturates \cite{Dias2018}. 

The frequency-dependent relative permittivities of Ag are taken from experimental data~\cite{Babar2015}. We consider $p$-polarized plane-wave impinging on the hybrid system and restrict our discussions to the normal incidence ($\vec{H}_i=H_i\hat{y}$ and $\vec{E}_i=-E_i\hat{x}$). For the lattices, we study a periodic structure where we calculate the absorption A, transmission T and reflection R of the system, while in the case of isolated silver nanostructures we consider a finite system with perfect matched layers (PML) and calculate the extinction cross-section.  

Fig.\ref{fig1}a illustrates the typical resonance spectra for an isolated rod, where we show the resonance of the first mode and several subsequent  higher order resonances. Fig.\ref{fig1}b and c show the real part of  $E_z(\vec{r})$ ($\Re E_z(\vec{r})$) in the vicinity of the rod for the first and the third modes, respectively, where it can be seen that the field is highly confined in the insulating layer. Laterally, the AGPs propagate back and forth between the edges of the rod, generating a Fabry-Perot resonator. These confined AGP modes have wavelengths $\lambda_{AGP}$ given by $\lambda_{AGP}=2w/n$ where $n$ is the mode number. 

 Fig.\ref{fig1}d illustrates the typical resonance spectra for the periodic grating, where we show the resonance of the first mode and several subsequent resonances for two different values of $L$ and the same lateral cross-section $W$. We first notice that the spectra is strongly dependent on $L$. The field profile of the first modes under the metallic nanostructures are very similar for both isolated and periodic systems, even though the systems are
conceptually different. This is because of the creation of a AGP Fabry-Perot resonator under the rod. However, this is not valid for the outside region. In the periodic system, there are also GPs propagating back and forth in the region between rods (with size $L-W$), forming another resonator with wavelength $\lambda_{GP}$. By inspecting Fig. \ref{fig1}e and f, one can notice that in the limit of large periods where $L\ge 2W$, the resonances of the system are a consequence of the coupling of these modes, highlighted by the wavelengths $\lambda_{GP}$ and $\lambda_{AGP}$. In particular, if $L-W=W$, as in the case of $L=150$ nm, the two standing waves always belong to the same group of harmonics and we observe that the absorption peaks decrease monotonically with frequency. In the more general case, the resonances are a complex combination of coupled Fabry-Perot AGP and GP modes and there are variations in the height of the absorption peaks \cite{Epstein2020}.

\begin{figure}[h]
\centering{\includegraphics[width=0.97\columnwidth,clip]{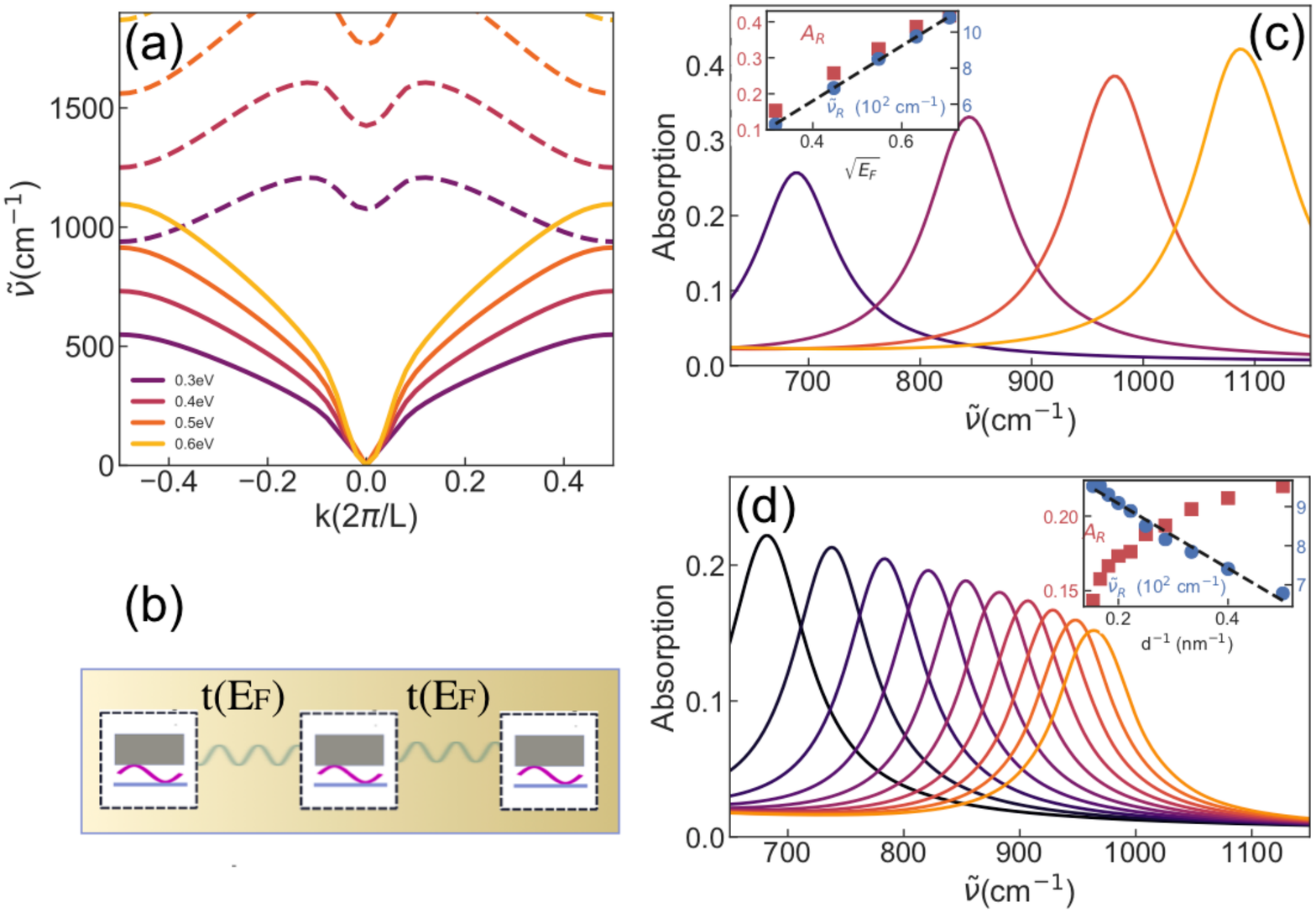}}
\caption{\label{fig2} (a) Band structure of  propagating AGPs for a system of periodic rods with square cross-section of lateral size $W=75$ nm. The solid lines represent the first band while the dashed lines represent the second band for increasing values of $E_F$ represented by an increase of the luminosity of the lines. (b) Schematic representation of the mechanism for the formation of a plasmonic band where there is an effective hopping between orbitals formed by the different modes of the AGP underneath the periodic metallic nanostructures.  (c) Absorption as function of the wave number $\tilde{\nu}$  for the same Fermi energy $E_F$ of panel (a), showing that the absorption spectra correspond to states at the edge of the band.  (d)  Absorption as function of spacer height $d$ (2 nm$\le d\le$6.5 nm) where increasing values of $d$ correspond to curves with increasing luminosity. The insets in (c) and (d) show the amplitude of the resonant absorptions $A_R$ (right) and their wave-number $\tilde{\nu}_R$  in function of the increasing value of the parameter. }
\end{figure}

To highlight these differences, we further explore the periodic system by calculating its eigenfrequencies as a function of the momentum. For simplicity, from now on we shall focus on systems with $W=75$ and $L=150$ nm where $L-W=W$. Fig. \ref{fig2}a show a band structure for the periodic rods where their bandwidth increases for increasing values of $E_F$. We can thus conclude that although isolated and periodic nanorods have similar field enhancement of the first order mode, they are essentially different plasmonic devices. Isolated rods give rise to localized strong electromagnetic fields only under the metallic structure. 

 This behavior can be understood in an intuitive picture as follows: depending on the dimensionality of the metallic nanoparticles, the MIG structure can be seen as a  quantum dot or quantum well and, in analogy with electronic waves, effective orbitals give rise to localized strong electromagnetic fields under the isolated metallic structure, while periodical structures give rise to propagating plasmons with highly modulated electromagnetic fields.  When organized in a periodical lattice of the order of hundreds of nanometers, the “effective electromagnetic orbitals” (coupled dipoles or quadrupoles) form bands of propagating AGPs, similarly to Bloch bands originated from localized orbitals in solids~\cite{Bylinkin2019}. Different from electrons in solids however, the bandwidths can be controlled by gating, as GPs can be tuned with $E_F$, changing its optical conductivity $\sigma_g(\omega)$. The hopping between orbitals, as illustrated in the sketch of Fig.\ref{fig2}b, is mediated by the standing-waves of GPs and consequently depends on  $E_F$ which  thus can be used to modify the bandwidth. 
 
 As expected for AGPs, the resonances wavenumbers $\tilde{\nu}_R$ scale with $\sqrt{E_F}$, as can be seen in Fig. \ref{fig2}c, and the values of $\tilde{\nu}_R$ are compatible with the wavenumbers of the edges of the bands of Fig. \ref{fig2}a. In the range of interest, the absorption also increases linearly with $\sqrt{E_F}$ and the peaks become sharper, indicating that high doping regimes are better suited to maximize the AGPs resonances and the absorption efficiency, which is important for technological applications.  

Experimentally, it is possible to provide a vast degree of control over the light-AGP interaction. Besides the use of gating, the spectral range and absorption efficiency of the AGP resonances can be further tuned by varying the distance $d$ between the graphene and metal, the spacer $\varepsilon_d$ and substrate $\varepsilon_S$ permittivities. The characteristics of  the metallic nanostructure, such as the periodicity $L$ and the lateral size of the rods $W$~\cite{Iranzo2018} can also tune the AGP resonances (see Sup. Inf.). The separation between graphene and the rods has a strong influence on the confinement of the field \cite{Iranzo2018}, translated in the height of the resonance peaks in Fig. \ref{fig2}c, that decrease linearly with $d^{-1}$ (see inset). Still, $d$ can also be used to tune the resonance frequency without compromising the absorption efficiency.  By decreasing the values of $d$ in the range of 2-10 nm, the resonances are red-shifted, as in the case of nanocubes on top of silver substrates~\cite{Lassiter2013}.

\begin{figure}[h]
\centering{\includegraphics[width=0.97\columnwidth,clip]{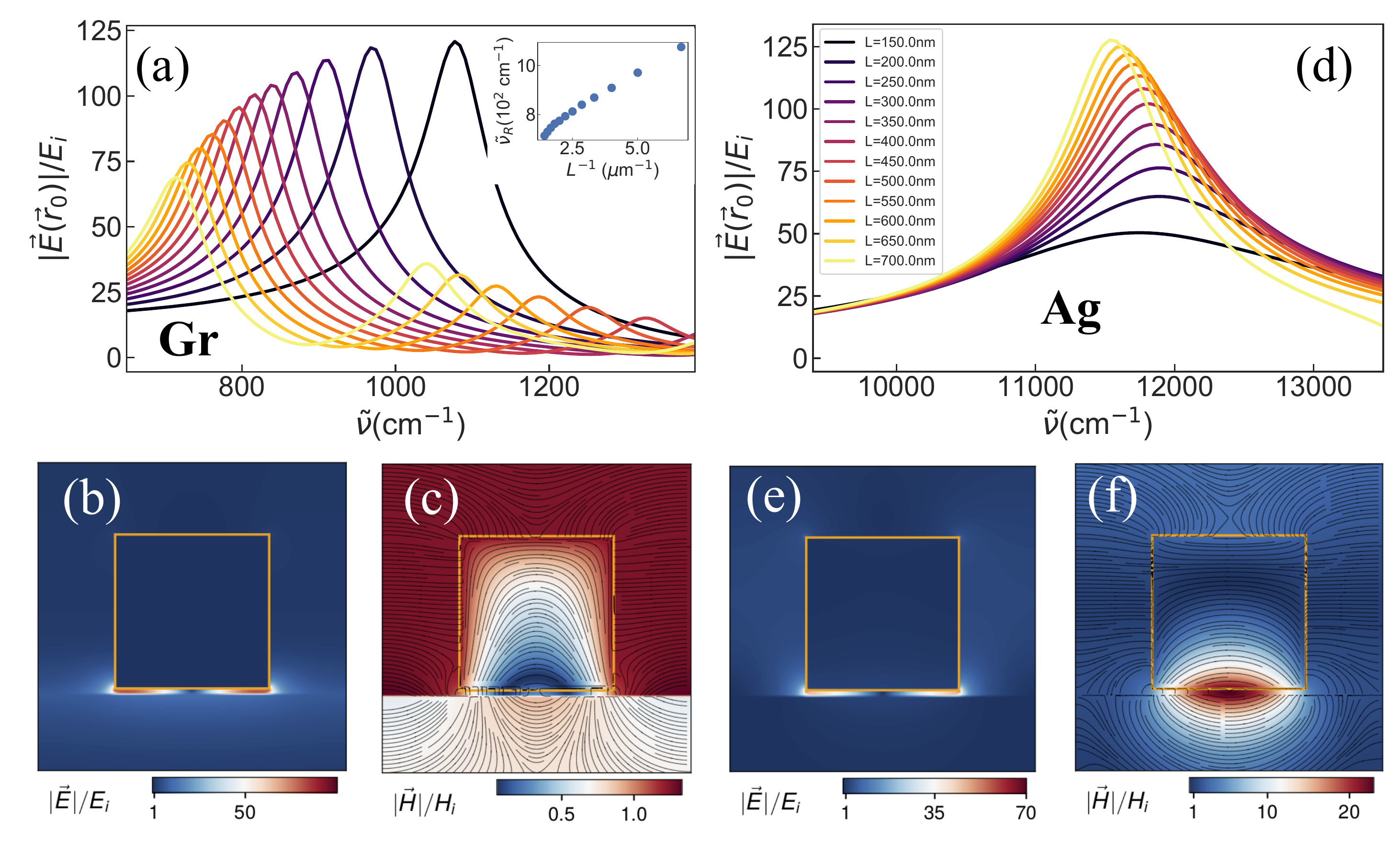}}
\caption{\label{fig3} (a) Electric field enhancement $|\vec{E}(\vec{r}_0)|/E_i$ of the main resonance mode at  $\vec{r}_0$, located at the edge of the rod, inside the spacer for the MIG setup and increasing values of the period $L$, $W$=75 nm and $E_F$=0.5 eV. The inset shows a linear dependence of the resonance wavenumber $\tilde{\nu}$ in function of $L$. (b) Electric field distribution $|\vec{E}(\vec{r})|/E_i$ and (c) Magnetic field distribution $|\vec{H}(\vec{r})|/H_i$ for the resonance in (a) with $L=150$ nm. The streamlines in panel (c) illustrate the $\vec{E}(\vec{r})$ in plane $xy$.  Panels (d)-(f) present an equivalent analysis for  the MIM setup.  }
\end{figure}

We now turn to the comparison between the above obtained results for AGPs, with the same analysis for SPPs on similar structures.  To compare both systems, we consider the electric field enhancement at  $\vec{r}_0$, located at the edge of the rod, inside the spacer.  Fig. \ref{fig3}a shows the AGP resonances for increasing values of $L$. One can see that the resonance frequency of MIG structures is widely tuned simply by varying the period of the structure and it scales with $L^{-1}$. In contrast, the substantial field enhancement is weakly dependent on the period, slightly reduced for increasing values of $L$. This is accompanied by a $L^{-1}$ dependency of the absorption, which is proporcional to the density of rods per unit length (see Sup. Inf.). These characteristics are in sharp contrast with the equivalent hybrids of MIM SPP systems (see Fig. \ref{fig3}d), where there is little variation of $\tilde\nu_R$ with L, a consequence of the small dispersion of SPP modes in MIM devices. However, contrary to what is observed for MIG SPP systems, the field enhancement of MIM devices is strongly dependent of $L$. The resonances become sharper for large $L$, leading to robust field enhancements that begin to saturate around $L=600$ nm (see Fig. \ref{fig3}e). This dependence with $L$ is similar to what is observed in plasmons in 1D and 2D arrays of metallic nanoparticles, where the resonance results from the interplay between the excitation of localized plasmons on the particles and the diffraction by the periodicity of the lattice~\cite{Abajo2007}.  The width of the resonances also show that the cavities' quality factor has opposite dependency with $L$ for MIG and MIM SPP systems. While MIG resonators have best quality factors for small $L$, the quality factor improves with $L$ for MIM resonators, as observed in collective resonances of metallic particles~\cite{Abajo2007}.

Figure \ref{fig3}b and c show the typical distribution of the electrical and magnetic fields of the AGP resonance in the vicinity of a single rod in a MIG structure, while panels Figure \ref{fig3}e and f provide the same distributions for MIM device.  Both systems have similar electric field distribution $\vec{E}(\vec{r})$ extremely confined to the spacer with a dipole like structure in $E_z(\vec{r})$. However, they have different magnetic field distributions. The 2D character of graphene plays an important role in the magnetic field distribution. The two systems have displacement currents in the $xz$ plane, generating the magnetic field inside the rod. But they run in opposite directions. While one enhances the incident field, the other suppresses it.

Many characteristics of plasmonic resonances in nanocubes are similar to those in rods~\cite{Wu2011}. However,  the geometry of the nanostructure has strong influence in the confinement of the electromagnetic field. Fig. \ref{fig3}b shows that the electric field  is strongly localized at the edges of the rods. However, because they are infinite in $y$, the field is delocalized in one dimension, forming lines of highly localized electric field. In this sense, rods can be seen as plasmonic waveguides. On the other hand, in a cube (a zero-dimensional system), the electric field is constraint in all three dimensions. Consequently, there are important differences in the field distributions of the two setups, having a strong influence in the magnetic field distribution. Furthermore, the dimensionality of the periodic structure should also have an influence on the resonance dependence with the size of the unit cell $L$. 

\begin{figure}[h]
\centering{\includegraphics[width=\columnwidth,clip]{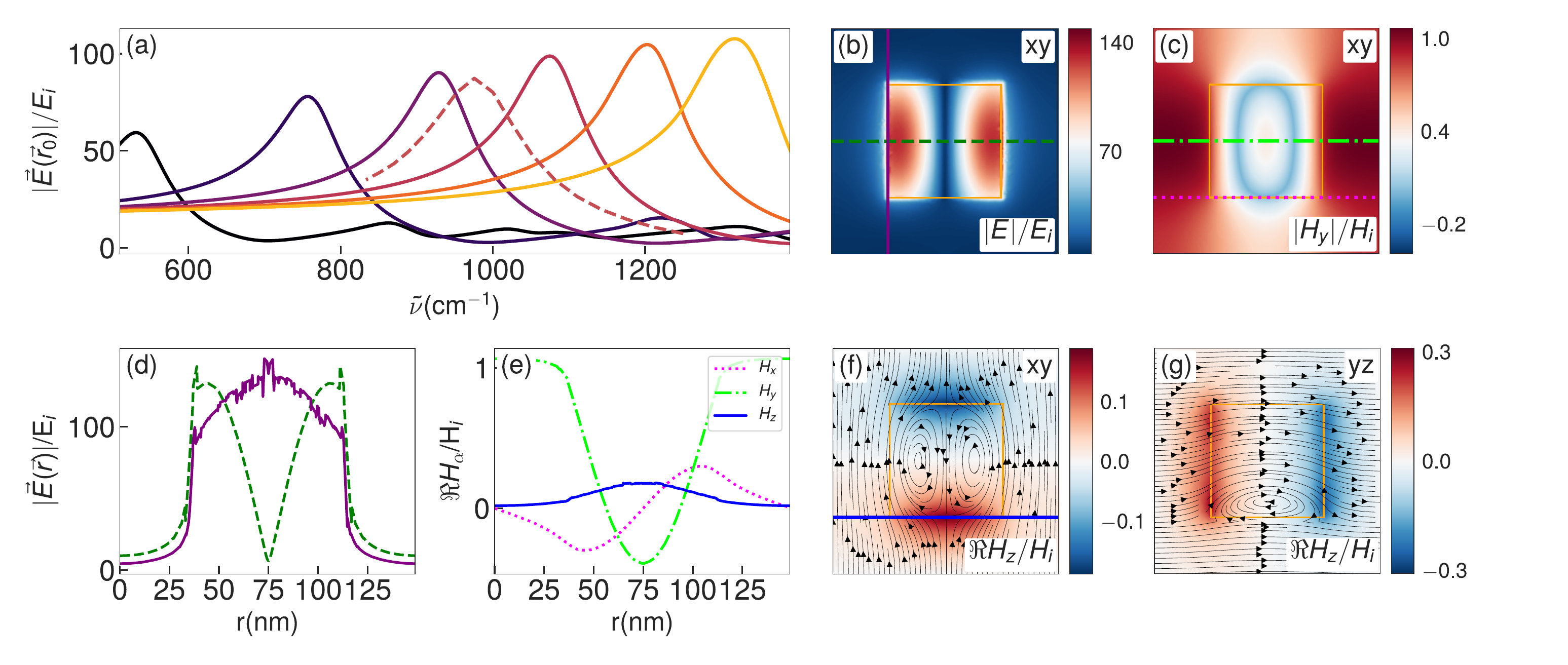}}
\caption{\label{fig4}  (a) Magnitude of the electric field $|\vec{E}(\vec{r}_0)|$ under a corner of the cube as a function of $\tilde{\nu}$ for increasing values of $E_F$, represented by an increase of the luminosity in the different curves. The dashed line represents $|\vec{E}(\vec{r})|$ for an isolated cube.   (b) and (c)  density plot of $|\vec{E}(\vec{r})|$ and  $| H_y(\vec{r})|$ respectively.  (d) $|\vec{E}(\vec{r})|$ in function of $\vec{r}$ for the two paths indicated in (b). (e) presents the real values of the three components of the magnetic field for the three paths indicated  in (c) and (f), that presents the density plot of $\Re H_z(\vec{r})$  while (g) shows the density plot of  $\Re H_z(\vec{r})$ in the plane $yz$ while   The streamlines in (f) represent the in-plane components of $\vec{H}(\vec{r})$  in the $xy$ plane while the streamlines in (g) represent the in-plane components of $\vec{H}(\vec{r})$  in the $yz$ plane. The intensities of the electric and magnetic fields are normalized by the intensity of incident fields $E_i$ and $H_i$ respectively.  } 
\end{figure}

To begin exploring the characteristics of the AGP resonance in nanocubes, let us first consider a periodic structure of nanocubes with $W=75$ nm and $L=2W$. Similar to the case of rods, the field enhancement $|\vec{E}(\vec{r})|/E_i$ underneath a cube is also of the order of 150. To illustrate the enhancement, we calculate $|\vec{E}(\vec{r}_0)|/E_i$ where $(\vec{r}_0)$ is located in the spacer underneath a corner of cube, at a distance of 2 nm from graphene. Figure \ref{fig4}a presents $|\vec{E}(\vec{r}_0)|/E_i$ for increasing values of $E_F$. The resonance frequency still scales linearly with $\sqrt{E_F}$  while the magnitude of the field saturates in the high doping regime. 

Different from the case of rods, where $\tilde{\nu}_R\propto 1/L$, for periodic cubes, the main peaks have minimal variation with $L$ for the values considered here (although high modes are sensitive to changes in the period). For the isolated cube, the field enhancement is similar to the periodic system. This can be inspected in Fig.\ref{fig4}a where the dashed line shows $|\vec{E}(\vec{r})|/E_i$  for a single cube in the same location of the periodic systems.

Let us now proceed to discuss the field distribution of the nanocubes. Similar to the rods, the field distributions of metallic nanocubes for the main resonance mode are qualitatively the same for a lattice for $L\ge 2W$ and for isolated cubes. For convenience, we present the results for a periodic system.  To analyze the difference between the field distributions of the rod and the cube, let us first focus on the $xy$ plane at a distance of 2 nm from graphene. Panel \ref{fig4}d shows $|\vec{E}(\vec{r})|/E_i$  for two different cuts of the density plots shown in panel \ref{fig4}b. The functional form of $|E(x)|$ is very similar to the rod but the purple curve indicates that the electric field also localizes in $y$ and its maximum is located at the two edges $x=\pm W/2$ of the cube for $y=W/2$. Similarly to what was seen in Figure \ref{fig1}b, $E_z(\vec{r})$ has opposite signs for $x=\pm W/2$. This resonance mode can also be understood as a waveguide mode like the ones for the rods, where the electric field distribution of the cavity mode satisfies the half-wavelength criteria. It is consistent with this interpretation which explains the resonance's frequency variation with the size of the cube following the simple formula $\lambda^{-1}_R\propto W^{-1}$. This linear dispersion relation is expected for AGPs~\cite{Epstein2020}: for increasing nanocube sizes, the length of the cavity increases and supports a resonance at a longer wavelength.  

The main differences between the localized electromagnetic field in the vicinity of rods and cubes can be seen by analyzing the magnetic field distribution. Qualitatively, the distribution of $H_y(\vec{r})$  in the cube is similar to the one of the rod. However, it has some important differences when compared with the one of Fig. \ref{fig3}f: The intensity of the magnetic field has a much more pronounced variation inside the cube. This variation is confined to a small semi-oval region located in the lower part of the cube and spacer (see Fig. \ref{fig4}g), giving rise to a magnetic nano-resonator. 

 $H_z(\vec{r})$ has a sizeable magnetic dipole contribution, seen in the density plot of Fig. \ref{fig4}f.  In the same panel, the streamline represents the in-plane contribution of the magnetic field, which is typical of a magnetic dipole. The values of  $H_x(\vec{r})$, $H_y(\vec{r})$ and $H_z(\vec{r})$ in the $xy-$plane at a distance of 2 nm from graphene are presented in Fig. \ref{fig4}e  for three different cuts of the density plots shown in panels c and f where it is clear that the minimum of the magnetic field occurs at the center of the cube and the $x$ and $z$ components have the same magnitude. Differently from the electric field, which is concentrated underneath the cube, $H_x(\vec{r})$ and $H_z(\vec{r})$ have large values in the laterals of the cube, as shown in Fig. \ref{fig4}g.  Still, the streamlines in Fig. \ref{fig4}g, presenting the $yz-$components of $\vec{H}(\vec{r})$, show that the magnetic dipole resonator is located at the lower part of the cube, exactly at the border of the dielectric spacer. 

Magnetic dipole nano-resonators are essential building blocks for the design of metamaterials with on-demands electric and magnetic response. For nano-resonators based on metallic surfaces, the plasmon wavelength is determined by the size of the nanostructures and characteristics of the setup.  However, MIG devices allow an in-situ variation of the resonance frequency by gate voltage.  The results above show that it is possible to modify the characteristics of the magnetic nano-resonators by a gate, which can be a form of controlling the electric and magnetic response of the metamaterial electrically.

In conclusion, we have shown that different metallic nanostructures graphene hybrids can localize AGPs leading to high field confinement and enhancement in various dimensions that can be tuned by gating. These characteristics, together with the strong field gradients, are ideal not only for sensing but also for nano-optical manipulation of small particles through confining potentials and optical forces~\cite{Zhang2016}.  Our results indicate that hybrids with periodically assembled nanostructures generate plasmonic crystals with gate-tuned band structures. The propagating AGPs give rise to a strong periodically modulated field enhancement that could be used as a versatile plasmon modulator. Furthermore, dispersed metallic nanocubes on top of graphene work as electrically controlled dipole magnetic resonators that could be used as building blocks for novel and  highly tunable metamaterials.

\begin{acknowledgements}
N.M.R.P. and F.H.L.K. acknowledge support from the European Commission through the project “Graphene-Driven Revolutions in ICT and Beyond” (Ref.  No. 881603, CORE 3). N.M.R.P. and T.G.R. acknowledge COMPETE 2020, PORTUGAL 2020, FEDER and the Portuguese Foundation for Science and Technology (FCT) through project POCI-01- 0145-FEDER-028114. F.H.L.K. acknowledges financial support from the Government of Catalonia trough the SGR grant, and from the Spanish Ministry of Economy and Competitiveness, through the “Severo Ochoa” Programme for Centres of Excellence in RD (SEV-2015- 0522), support by Fundacio Cellex Barcelona, Generalitat de Catalunya through the CERCA program, and the Mineco grants Ramón y Cajal (RYC-2012-12281, Plan Nacional (FIS2013-47161-P and FIS2014-59639- JIN) and the Agency for Management of University and Research Grants (AGAUR) 2017 SGR 1656.  This work was supported by the ERC TOPONANOP under grant agreement n 726001 and the MINECO Plan Nacional Grant 2D-NANOTOP under reference no FIS2016-81044-P.
\end{acknowledgements}

\section*{Authors' contribution}
T. G. R. and I. E contributed equally to this work.

%

\end{document}